\documentstyle[12pt,twoside]{article}
\renewcommand{\baselinestretch}{1.4}
\input epsf
\newcommand{\beqna}{\begin{eqnarray}}
\newcommand{\eeqna}{\end{eqnarray}}
\newcommand{\beqn}{\begin{equation}}
\newcommand{\eeqn}{\end{equation}}

\newcommand{\plabel}[1]{\label{#1}}
\newcommand{\pbibitem}[1]{\bibitem{#1}}

\begin{document}

\title{\small \rm \begin{flushright} \small{hep-ph/9701425}\\
\small{RAL-97-002}\\
\small{MC-TH-96/31}  \\
\end{flushright} \vspace{2cm}  
\LARGE \bf Distinguishing Hybrids from Radial 
Quarkonia \vspace{0.8cm} }
\author{Frank E. Close \thanks{E-mail : fec@v2.rl.ac.uk} \\
{\small \em Particle Theory, Rutherford--Appleton Laboratory, Chilton,
Didcot OX11 0QX, U.K.} \\
Philip R. Page \thanks{E-mail : prp@jlab.org}
\thanks{\small \em Present address:
Theory Group, Thomas Jefferson National Accelerator Facility, 
12000 Jefferson Avenue, Newport News, VA 23606, U.S.A.
} \\
{\small \em Department of Physics and Astronomy, University of Manchester}\\
{\small \em Manchester M13 9PL, U.K.}
 }
\date{January 1997}
\vspace{1.5cm}

\begin{center}
\maketitle

\begin{abstract}
 
We present arguments
that reinforce the hybrid interpretation of $\pi(1800)$ and we 
establish that the
$\rho(1450)$ and the $\omega(1420)$ can be interpreted as radial--hybrid mixtures.
Some questions for future experiments are raised.
\end{abstract}
\end{center}

\newpage
 Evidence for the excitation of gluonic degrees of freedom in strong QCD
has recently emerged with the possible discovery of 
a hybrid with $J^{PC} = 0^{-+}$ \cite{ves,bussetti,rya,ves951,ves96} 
in the $1.8$ GeV mass region. Both its mass and unusual decay
patterns are as expected for such gluonic excitations \cite{paton85,bcs}. 
Idiosyncratic decay patterns have also been noted for $1^{--}$
in the $1.4 - 1.7$ GeV region \cite{cd1,kalash}. These are in line with the predictions of the extensive study of 
ref. \cite{cp95} for hybrids with non--exotic overall 
$J^{PC} = 0^{-+}, 1^{--}$. 

If these states are not hybrids then radially excited
quarkonia are the only conservative alternative.
In ref. \cite{page96rad} the point of departure was to perform a ``control"
test by attempting to assign these states to be radial excitations
of conventional quarkonia, compute the expected branching ratios for these
radial states following the standard prescriptions of 
refs. \cite{oliver,kokoski87} and then compare the data against these as well as the
gluonic hypothesis. The analysis concluded that hybrid excitations
appear to be manifested in the data.

This is a radical result if true and merits critical examination.
Here we test its robustness by seeking to relax some implicit
assumptions.  In ref. \cite{page96rad} the results were all in the 
special case where the wave function parameter $\beta_A$ of the incoming state 
is the same as $\beta$ of the outgoing states.
In the present paper we relax\footnote{All calculations have been done in the conventions
of refs. \protect\cite{cp95,kokoski87}, which differ in phase space
vconvention and overall normalizing constant from
ref. \protect\cite{page96rad}. The normalizing constant is fixed.} 
this by allowing $\beta_A$ to be different from $\beta$, i.e. to be ``off the
iso--$\beta$ axis''.
For this purpose, we use a ``standard parameter region'' where $\beta_A = 0.25 - 0.45$ GeV
and $\beta = 0.3 - 0.5$ GeV. 

The mass of $1^{--}\rho(1450)$ \cite{cd1,pdg96} suggests a natural assignment as
$2\: ^3S_1\;q\bar{q}$ \cite{godfrey} whereas its decays favour a hybrid interpretation
\cite{cd1,kalash,cp95}. By relying on the data analysis of ref. \cite{kalash} we 
are able to make stonger statements than ref. \cite{cp95} about mixing in the 
$1^{--}$ sector. In ref. \cite{page96rad} we argued 
 that a pure $2\: ^3S_1$ interpretation of $\rho(1450)$ is
untenable since its $\pi a_1$ and $\pi h_1$ modes cannot be simultaneously
accommodated. In the present paper we show that this conclusion
remains true even off the iso--$\beta$ axis. 
Moreover, we shall present arguments that $^3 D_1$ components
in $\rho(1450)$, $\omega(1420)$ and $\omega(1600)$ are insignificant, 
leaving us with a picture of hybrid--$2S$ mixing. 
The constitution of the $\rho(1700)$ is presently undetermined.
We highlight some channels where study at DA$\Phi$NE may illuminate these
questions further. These are discussed in section 1.

In section 2 we provide further arguments supporting the
hybrid interpretation of $0^{-+}\;\pi(1800)$, as proposed in 
Refs. \cite{cp95,page96rad,page96panic}. 

\section{$2S$ Radialogy: $2\: ^3S_1$ $\rho$ and $\omega$}

Given the masses of the $2\: ^1S_0$ states around $1.3$ GeV and that
the hyperfine
splitting in S--states tends to elevate the masses of the $2\: ^3S_1$
members of the supermultiplet, it is natural on mass alone to assign the
$\rho(1450)$ \cite{cd1,pdg96} and the $\omega(1420)$ to the $2\: ^3S_1$
levels of the spectrum \cite{godfrey}. Furthermore, they are some 300 MeV
below the predicted $^3D_1$ states 
which in absence of mixing are expected around 1.7 GeV,
and also lighter than unmixed hybrids which are predicted at $1.8 - 1.9$ GeV \cite{paton85,bcs}.
However, it is possibile that spin dependent forces may lower the mass of the
hybrid $\rho$ and $\omega$
(which are spin $S=0$ in contrast to the conventional 
$q\bar{q}$ components which are $S=1$) and cause
mixing between hybrid and conventional quarkonia. Thus 
one should {\it a priori} allow in
this region for the possibility of a triplication of states

\begin{equation}
|V\rangle \equiv cos\phi ( cos\theta |2\: ^3S_1\rangle + sin \theta |^3D_1\rangle)
+ sin\phi |V_H \rangle \plabel{trip}
\end{equation}

\renewcommand{\baselinestretch}{1}
\begin{table}
\caption{Widths of selected decay modes of radial $2\: ^3S_1$.
A range of widths in MeV is indicated on the iso--$\beta$ axis from 
$\beta_A,\:\beta = 0.3$ to $\beta_A,\:\beta = 0.45$, 
and in a band of thickness 0.15 GeV
around the iso--$\beta$ axis. The direction in which the width
increases is indicated along the
iso--$\beta$
axis and perpendicular to the iso--$\beta$
axis (under ``band'') ,
using the axis conventions of Fig. 3.
The number of nodal lines crossing the standard parameter region 
is also indicated.}
\plabel{tab1}
\begin{center}
\begin{tabular}{|c|c||r|r|c|}
\hline %------------------------
State & Mode & Iso-$\beta$ & Band & Nodal Lines\\
\hline \hline %------------------------
$\rho(1450)$ & $\pi\pi$ & 10 - 90  $\;\nearrow$ & 5 - 110  $\;\nwarrow$ &  1 \\ 
             & $\omega\pi$ & 90 - 120   $\;\nearrow$  & 50 - 160   $\;\nwarrow$ &  1 \\
             & $\rho\eta$ & 20 - 30  $\;\swarrow$ &  20 - 40  $\;\nwarrow$ &  1 \\
             & $K\bar{K}$ & 60 \hspace{0.54cm} & 30 - 90  $\;\nwarrow$ &  1 \\
             & $K^* \bar{K}$ & 20 - 40  $\;\swarrow$ & 20 - 40  $\;\nwarrow$ &  1 \\
             & $\pi a_1$ & 5 - 10  $\;\nearrow$ & 5 - 80 \hspace{0.54cm}   &  0 \\
             & $\pi h_1$ & 5   $\;\nearrow$ & 5 - 30 \hspace{0.54cm}  &  0 \\
\hline 
$\rho(1730)$ & $\pi\pi$ & 1 - 80   $\;\nearrow$ & 1 - 100  $\;\nwarrow$ &  1 \\  
             & $\omega\pi$ & 40 - 170   $\;\nearrow$  & 20 - 220  $\;\searrow$ &  1 \\
             & $\pi a_1$ & 20 - 50   $\;\swarrow$& 20 - 110  \hspace{0.54cm} &  0 \\
             & $\pi h_1$ & 30 - 70   $\;\swarrow$ & 30 - 70  \hspace{0.54cm}  &  0 \\
\hline \hline %------------------------
$\omega(1420)$ & $\rho\pi$ & 270 - 350   $\;\nearrow$ & 160 - 450  $\;\searrow$ &  0 \\
               & $\pi b_1$  & 5  $\;\nearrow$  & 5 - 40  \hspace{0.54cm} &  0 \\
\hline 
$\omega(1600)$ & $\rho\pi$ & 190 - 480   $\;\nearrow$ & 110 - 620  $\;\searrow$&  1 \\
               & $\pi b_1$ & 20 - 40  $\;\swarrow$ & 20 - 100 \hspace{0.54cm}  &  0 \\
\hline %------------------------------
\end{tabular}
\end{center}
\end{table}
\renewcommand{\baselinestretch}{1.4}

A well known problem for the radial assignment of $\rho(1450)$,
 ($\phi,\theta \rightarrow 0$),
 is that the relative partial widths
of the state appear idiosyncratic \cite{cd1,kalash,cp95}.  The signals
appear to be in remarkable agreement with those predicted for a hybrid
($\phi \rightarrow \frac{\pi}{2}$)  \cite{cp95,page96rad}.
These are very different from the historical predictions of radial or $^3D_1$
decays of quarkonia \cite{page96rad,oliver,kokoski87}.
In particular
the the experimentally observed suppression \cite{cd1} of $\pi h_1$ relative to $\pi a_1$ is, within the flux--tube
model, a crucial test of the hybrid initial state. This empirical
 result contrasts with the behaviour expected 
of a $^3D_1$ for which both $\pi h_1$
and $\pi a_1$ are predicted to be large \cite{page96rad,oliver,kokoski87} and also with
the case of the $2\: ^3S_1$ where both of these channels are predicted
 to be small. 
Some partial widths for a $2\: ^3S_1$ initial state are shown in Table
\ref{tab1}. The reason for the suppression of $\pi h_1$ in hybrid
$1^{--}$ decays is because in the hybrid
the $q\bar{q}$ has $S=0$, whereas for the ``conventional
quarkonium" $1^{--}$ the $q\bar{q}$ have $S=1$; the $^3P_0$ decay is forbidden
by spin orthogonality in the former example for final states where the
mesons' $q\bar{q}$ have $S=0$, as in the $\pi h_1$ case.
 It is therefore
interesting that the detailed analyses of refs. \cite{cd1,kalash} commented
on the apparently anomalous decays that they found for
the $1^{--}$ state $\rho(1450)$, in particular the suppression of $\pi
h_1$ relative to a prominent $ \pi a_1$; specifically 
\beqn
\plabel{rhodata}
\begin{array}{cccccc}
\pi a_1 + \rho(\pi\pi)_S & \pi h_1  +  \rho \rho + \rho(\pi\pi)_S &
\omega\pi & \pi \pi & \eta\pi\pi &    \\ 
\; 190 & 0-39 & 50-80 & 17 -25 & 4-19\; & MeV \end{array}
\eeqn
There is no $2\: ^3 S_1$ solution consistent with the above 
data \cite{cd1}. The noticeable feature in the data is the
strong coupling to $\pi a_1$ relative to $\pi h_1$ which is greater than
$\frac{190}{40}$.
Ref. \cite{page96rad} noted that this is outside any sensible solution for a radial and so
 $\rho(1450)$ cannot be pure $2\: ^3S_{1}$. 

The stability 
of these conclusions with respect to independent variations in 
$\beta_A,\;\beta$ has not hitherto been assessed. This is the point of
departure for  the present paper.
To test the robustness of this 
conclusion we have studied what happens if we depart form the ``iso-$\beta$"
contour in $\beta$ space and allow the initial and final values to differ.
Fig. \ref{2srhophil} shows that the $\pi a_1$ and $\pi h_1$ widths
form a valley in $\beta$ space. We can climb the
valley walls to elevate the $\pi a_1$ rate but this elevates $\pi h_1$ too,
contrary to experiment where $\pi h_1 < 40$ MeV $\approx \frac {1}{5} 
\pi a_1$.

Thus the conclusions are robust if present data are reliable. 
If the experimental rate of $\pi a_1$ were reduced by 50\%
then it could be possible to describe the state as $2\: ^3S_1$ with $\beta_A = 0.35$ GeV,
$\beta = 0.4$ GeV for which 
\begin{equation}
\pi a_1 : \pi h_1 : \omega\pi : \pi \pi = 75 : 25 : 75 : 25 \; MeV
\end{equation}
though there is no experimental indication of reduced $\pi a_1$. If instead
one accepts the $\pi a_1,~ \omega\pi$ and $\pi \pi$ data, but ignores
$\pi h_1$, there is the following possibility for $2~^3S_1$ with
$\beta_A = 0.4$ GeV, $\beta = 0.5$ GeV
\begin{equation}
\pi a_1 : \pi h_1 : \omega\pi : \pi \pi = 165 : 50 : 45 : 25 \; MeV
\end{equation}
This highlights the importance of quantifying the $\pi h_1$ channel
with new data, in particular
in dedicated $e^+ e^-$ experiments.

Now we turn to the $\omega(1420)$ and $\omega(1600)$ pair. The first
inference is that neither can have a significant $^3D_1$ component. 
The $\omega(1420)$ data have $\pi b_1 \sim 0$ MeV \cite{cd1}. 
The $\omega(1600) \rightarrow \pi b_1$ also is small
($\sim 30$ MeV) \cite{cd1}.
If these data are confirmed it would rule out $^3D_1$ 
($\theta \sim \frac{\pi}{2}$) for the $\omega(1420)$ and also for the
$\omega(1600)$ as $\pi b_1$ is predicted to dominate the $^3D_1$ decays
in the iso--$\beta$ case \cite{page96rad}. The effect of relaxing the iso--$\beta$ constraint is
illustrated in Fig. 2 for the $\omega(1600)$ (results for $\omega(1420)$ are
similar). For most of the parameter space the width exceeds 100 MeV and
nowhere falls below 30 MeV which reinforces the conclusion that
$^{3}D_1$ is incompatible for these states. 

Having eliminated $^3D_1$, then within the three
state mixing hypothesis of Eq. \ref{trip} this leaves $2\: ^3S_1$ and
hybrid as possible configurations. Either
of these is consistent with the $\pi b_1$  channel being small: (i) for
the hybrid, the spin selection predicts $\pi b_1$ to vanish; (ii) 
the $2\: ^3S_1$ ($\theta \sim 0$) has $\pi b_1 \sim 5$ MeV 
for $\omega(1420)$ and $\sim 30$ MeV for $\omega(1600)$ on the
iso--$\beta$ axis. 
In addition, for radials, $\Gamma(\omega(1600) \rightarrow 
b_1\pi)$ $\geq$ $2 \; \Gamma(\omega(1420) \rightarrow b_1\pi)$ in
the standard parameter region, consistent with the data \cite{cd1}. 

Within the $2\: ^3S_1$--hybrid space, data are incompatible with
$2\: ^3S_1$ alone. If $\omega(1420)$ were pure $2\: ^3S_1$, this small value for $\pi b_1$
would imply that its $2\: ^3S_1\;\rho$ partner would also have a small
$\pi a_1$ width for the same $\beta$'s. Thus if the $\rho(1450)$ and $\omega(1420)$ have
similar internal structure then $\omega(1420)$ cannot be pure $2\: ^3S_1$.
The $e^{+}e^{-}$ widths of
$\omega(1420)$ and $\omega(1600)$ are almost the same \cite{cd1},
which suggests strong $2\: ^3S_1 - V_H$ mixing. Thus

\begin{figure}
\hspace{2.6cm} \vspace{-2.3cm}
\hbox{\epsfxsize=5in}
\epsfbox{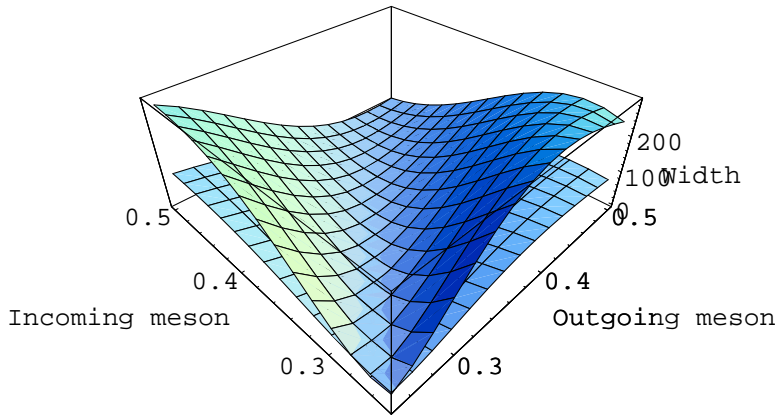}
\caption{Total widths in MeV of $2\: ^3 S_1 \; \rho(1450)\rightarrow
a_1\pi,\; h_1\pi$ ($a_1\pi$ is the larger channel, i.e. the upper of the
two sheets), as a function of
 $\beta_A$ of the incoming and $\beta$ of the outgoing mesons in GeV.}
\plabel{2srhophil}
\end{figure}

\begin{equation}
\omega(1420;1600) = cos\phi |2\: ^3S_{1}\rangle + sin\phi |\omega_{H}\rangle
\end{equation}
%with $\phi \sim \frac {\pi}{4}$.
Note also that departure from the
iso--$\beta$ valley would destroy the $\Gamma(\omega \rightarrow \pi b_1) \sim 0$ MeV
result.
This implies that one cannot fit the small $\pi b_1$ width for both
$\omega(1420)$ and $\omega(1600)$ within a $2\: ^3S_1 - ^3D_1$ basis
alone even off the iso--$\beta$ valley, and reinforces the need for a 
hybrid component.

\begin{figure}
\hspace{2.6cm} \vspace{-2cm}
\hbox{\epsfxsize=5in}
\epsfbox{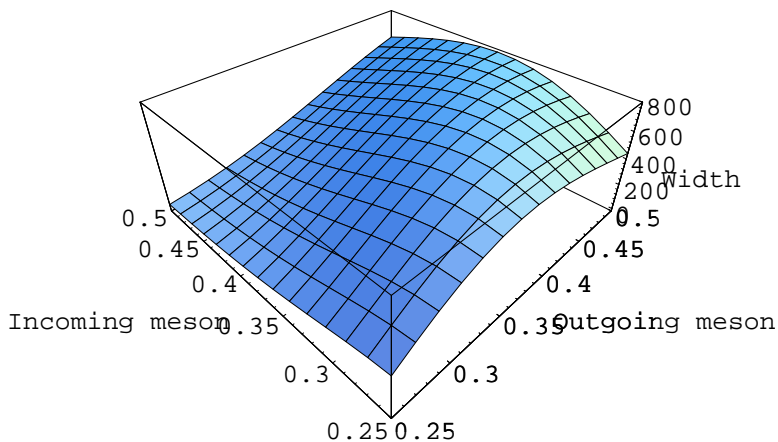}
\caption{Total widths in MeV of $^3 D_1 \; \omega(1600)\rightarrow
b_1\pi$, as a function of 
$\beta_A$ of the incoming and $\beta$ of the outgoing mesons in GeV.}
\end{figure}

The $\rho\pi$ decays are also consistent with $2\: ^3S_1 - V_H$ mixing.
For $\theta,\phi \rightarrow 0$, the channel $ \rho\pi$ dominates
with a predicted $2\: ^3 S_1$ width $\sim 350$ MeV for $\omega(1420)$
and $\sim 450$ MeV for $\omega(1600)$, which can become smaller
away from the iso--$\beta$ axis.
Experimentally $\Gamma (\omega(1420) \rightarrow
\rho\pi ) \sim 240$ MeV and for the $\omega(1600)$ the $ \rho\pi$  channel
is $85$ MeV \cite{cd1}; these results suggest a possible mixing
with a component that is ``inert" in the channel $\rho\pi $ such as the 
hybrid \cite{cp95}. 

This scenario of $2\: ^3S_1 - V_H$ mixing 
is also favoured by the $\rho(1450)$. A $2\: ^3S_1$ produces the 
$ \omega\pi$ as dominant mode (Table \ref{tab1}) and $\frac
{ \omega\pi}{\pi \pi} \sim 2-3$ for $\beta_A,\;\beta = 0.35 - 0.4$ GeV, 
results which are in accord with data
(Eq. \ref{rhodata}).
 For a hybrid the $\omega\pi $ is suppressed and
the $\pi \pi$ is zero.
 The presence of the $\pi \pi$ and $ \omega\pi$ channels hence calls for a $2\: ^3S_1$
component. 
However, $\rho\eta$ appears to favour hybrid, since the experimental signal is
very small (see Eq. \ref{rhodata} and E852 data \cite{manak}) and 
$2\: ^3S_1$ should have $\rho\eta$ at a strength of $\sim 30$ MeV. 
Hence a $2\: ^3S_{1} -V_H$ mixture is
a solution. Thus, as in the case of $\omega$, one has

\begin{equation}
\rho(1450) = cos\phi^{'} |2\: ^3S_{1}\rangle + sin\phi^{'} |\rho_{H}\rangle
\end{equation}
and the data can be driven by $\rho_H \rightarrow \pi a_1$ and 
$2\: ^3S_1\; \rho \rightarrow \pi \pi$. 

For $\rho(1700)$ the data indicate a very small $\omega\pi$ 
mode \cite{cd1}, pointing to
hybrid admixture, since $2\: ^3S_{1}$ and $^3D_{1}$ do not vanish, at
least along the iso--$\beta$ axis \cite{page96rad}. In order to force
vanishing one would need to move far off 
the iso--$\beta$ axis (see Table \ref{tab1}). 
However, the experimental $\pi\pi$ coupling of $\sim 100$ MeV
is substantial.
This is too large even for pure $2\: ^3S_1$ and $^3D_1$ at least in the
iso--$\beta$ limit, and certainly out of line with pure hybrid for
which this mode would vanish. If the experimental data survive there would be
a conundrum in that the small $\omega\pi$
and large $\pi\pi$ widths point in mutually incompatible
directions, namely the $\omega\pi$ favours hybrid while the 
$\pi\pi$ prefers radial $q\bar{q}$.
Errors in the experimental analysis can reduce the $\pi\pi$
coupling by up to 50\% \cite{privatedon}. 
Furthermore, a recent re--analysis of 
CERN--Munich data found a $\pi\pi$ width of only 
$39\pm 4$ MeV \cite{sarantsev96}. The true strength of the $\pi \pi$
coupling needs to be established.

The $\rho(1700)$
overall does not provide a strong constraint on our analysis. Within
the
large uncertainties the above are consistent with it being a $2\: ^3S_{1}
-V_H$ mixture
but do not demand it. Improved data in this region, such as at the
$e^+e^-$ facility DA$\Phi$NE, could be most useful: Specific channels 
that should be studied include $e^+e^-\rightarrow 4\pi$ in order to
separate $\pi h_1$ and $\pi a_1$ in the $4\pi$ state. New data in $\pi^+\pi^-\pi^+\pi^-$ have come from H1
at HERA \cite{hera}, and a coupled channel analysis is in progress at
Crystal Barrel \cite{meyerprivate}. Good data
on $\omega\pi$ and $\pi\pi$ are also needed.
  
Note that our scenario requires three $\rho$ (and three $\omega$) states
which should be allowed for in future data analyses. 

%***rho and omega consistent with the 1450 and 1700 being S--hybrid
%mixes and the D state suppressed also would solve the problem that
%otherwise there must be a third vector 
%meson; can we suppress production in e+e-? Yes if we say there is no D state 
%seen. Are you happy with this scenario? Analogous to $\psi(4040) -\psi(4160)$
%$3S$-hybrid; here it is $2S$-hybrid. Maybe because radial excitation gap 
%bigger for light quarks? 

%SUMMARY OF RECENT INFORMATION ON EXCITED RHO NOT CONTAINED IN OUR PAPER, BUT MENTIONED
%AT PANIC96:

%(1) C.A. Meyer (CBAR) reported rho(1450) $-> rho f_0 (1300)$ has been observed.

%(2) J. Manak (BNL) reported rho(1700) $->$ rho eta has been observed with
%rho(1700) parameters mass = 1.68 GeV, width = 287 MeV

%(3) Eugenio (BNL) reported two omega peaks in omega rho.

%Dr.  Curtis A Meyer: We are attempting an overall coupled channel
%analysis between pi+pi-3pi0 , pi+2p-2pi0, pi-4pi0 and 5pi0. None of this
%is written up at the moment, though we seem to see evidence
%for rho(1450) $->$ rho sigma, rho(1700) $->$ a1 pi, but not the other way.***

%The fact that $4\pi + 6\pi$ is found to be $\sim 300$ MeV \cite{cd1} 
%is also fully consistent
%with our finding that radial 
%$\rho(1700)\rightarrow a_1\pi > 1.3\: \rho(1450)\rightarrow a_1\pi$
%CHECK HYBRID] and radial 
%\rho(1700)\rightarrow h_1\pi > 1.5\; \rho(1450)\rightarrow h_1\pi$ in the
%standard parameter region.

%$2\: ^3S_{1}$ widths only reach 100 MeV off the iso-$\beta$ axis       
%and $^3D_{1}$ only reaches 100 MeV on the iso-$\beta$ axis at $\beta_A,\beta \sim 0.45$.

\section{$3S$ Radialogy: $ 3\: ^1S_0\; \pi$}

There is a  resonance $\pi(1800)$ in $\pi f_0(980)$,
$\pi f_0 (1300)$ and
also $ (K \bar{K} \pi)_S$. It is a common feature that $\pi(1800)$ is absent
in $\rho\pi $ and $K^*\bar{K}$. The presence of clear signals
in both $\pi f_0(1300)$ and $\pi f_0(980)$ is remarkable and was commented 
upon with some surprise \cite{ves}. A substantial branching ratio to
$\pi f_0(1500)$ has also been reported \cite{ves951,page96jpsi}.

\renewcommand{\baselinestretch}{1}
\begin{table}
\begin{center}
\caption{Widths of selected decay modes of radial $3\: ^1S_0\; \pi(1800)$.
Conventions are as in Table \protect\ref{tab1}.
For the mode $ K^*_0(1430)\bar{K}$ widths are indicated for a state at 2 GeV.}
\plabel{tab2}
\begin{tabular}{|c||r|r|c|}
\hline %------------------------
Mode & Iso-$\beta$ & Band &  Nodal Lines\\
\hline \hline %------------------------
$\rho\pi$ & 0 - 30  \hspace{0.54cm} & 0 - 70  \hspace{0.54cm} &  2 \\
$K^* \bar{K}$ & 30 - 50 $\;\nearrow$  & 5 - 110 $\;\searrow$ & 2 \\
$\rho\omega$ & 20 - 50 $\;\nearrow$  & 5 - 90 $\;\searrow$  &  2 \\
$K^* \bar{K}^*$ & 5 $\;\swarrow$  & 1 - 10 $\;\nwarrow$  &  2 \\
$\pi f_0(1300)$ & 0 - 5  $\;\swarrow$   & 0 - 5 \hspace{0.54cm}  &  2 \\
$\pi f_2(1270)$ & 10 - 20 $\;\swarrow$  & 10 - 30 $\;\nwarrow$  &  1 \\
$K^*_0(1430)\bar{K} $ & 5 - 10 $\;\nearrow$  & 0 - 10 \hspace{0.54cm}  &  2 \\
\hline %------------------------------
\end{tabular}
\end{center}
\end{table}
\renewcommand{\baselinestretch}{1.4}

In refs. \cite{cp95,page96rad,page96panic} $\pi(1800)$ has been argued to be
a hybrid meson. The overall expectations for hybrid $0^{-+}$ are in line with
the data of refs. \cite{ves,rya,ves951,ves96}, except that the signal seen in $\rho\omega$ and $\pi f_2$ 
might be a manifestation of $3\: ^1S_0$. 
In order to settle this question, it is imperative to compare the data
to the predictions for radial $3S$. Since $\rho\pi$
and $K^* K$ are experimentally \cite{ves} found to be suppressed, it is 
of significant interest whether this can also happen for radial. This
was discussed in the iso--$\beta$ case in ref. \cite{page96rad};
here in Fig. \ref{3snodes} we show the result of allowing $\beta_A\neq\beta$. 
We clearly see that there are  
``nodal lines in the amplitude'' for each of $\rho\pi$
and $K^* K$, by which we mean that the
amplitude as a function of $\beta_A$ and $\beta$ displays lines along
which the amplitude vanishes. 
Moreover, the same happens for $\rho\omega$ and $K^* K^*$.
For $\rho\pi$ the amplitude can vanish even on the iso--$\beta$ axis.
We conclude that radial decays to pairs of S--wave mesons can be
forced to 
vanish, although only in the case of the $\rho\pi$ channel does
this happen near to the iso-$\beta$ axis. 

The $\pi f_0(1300)$ is very much suppressed throughout the entire parameter space (see Table 
\ref{tab2}), relative to
the prediction for hybrid of 170 MeV. The same is true of $K^*_0(1430)\bar{K} $ which is small for
a $3\: ^1S_0\; q\bar{q}$, but large in the data (manifested as $(K\bar{K}\pi)_S$) and the
 largest channel for a hybrid $\pi_H$. This is most easily seen for states at
2 GeV, so that enough phase space for the decay to $ K^*_0(1430)\bar{K}$ is available.
For radial we have small widths due to nodal lines in the amplitude (see Table 
\ref{tab2}), 
while in contrast for hybrid the width is predicted to be 200 MeV.

Nonetheless, in this mass region we also expect the $3\: ^1S_0\; \pi$ to
appear and we now seek possible signatures. For a 
$3\: ^1S_0$ the $\rho \omega$ channel is expected to be prominent \cite{page96rad}. Fig. 3 shows
that the regions in $\beta$ space where $\rho \omega$ modes could be suppressed by
nodes are far from the physically favoured region and so we expect 
that $3\: ^1S_0 \rightarrow \rho\omega$ is indeed a prominent mode.
Note that this channel vanishes for hybrid and so the $\rho\omega$
channel promises to be a sharp discriminant between hybrid $\pi$ and
$3\: ^1S_0$ initial states. 
The $\rho \omega$ signal 
builds up significantly below 1800 MeV and also shows a high mass
continuum which looks somewhat different to the $\pi_H(1800)$. 
A resonant signal however has not yet been established, although
a ``resonance--like structure'' with mass $1742\pm 12\pm 10$ MeV and
width $226\pm 14\pm 20$ MeV has been reported \cite{ves96}.
The $\pi f_2$ channel also may
discriminate $\pi_H$ from $3\: ^1S_0$. For $\pi_H$ this is predicted to be
a minor mode whereas for $3\: ^1S_0$ it is predicted to be a more 
significant signal. Fig. 4d in ref. \cite{ves951} shows a clear
$\pi f_2$ peak at 1700 MeV, certainly
below the 1800 MeV region of the $\pi(1800)$ as already noted in 
ref. \cite{page96rad}. Further analysis and data
are now required to establish this. For hybrid $\pi f_2$ is 6 MeV \cite{page96rad}
while for radial it is possibly larger (see Table 
\ref{tab2}). It is tempting  to suggest that the 
$3\: ^1S_0$ favoured $\rho \omega$ and $\pi f_2$ channels peak at $\sim 1700$
MeV in contrast to the $\pi_H$ channel $\pi f_0$ at $\sim 1800$ MeV. If two
$0^{-+}$ states were to be isolated in this region this would be strong
evidence for hybrid and $3S$ excitation. Categorisation of $5\pi$/$3\pi$ may further clarify this possibility.

\begin{figure}
\hspace{2.6cm} \vspace{-.4cm}
\hbox{\epsfxsize=6 in}
\epsfbox{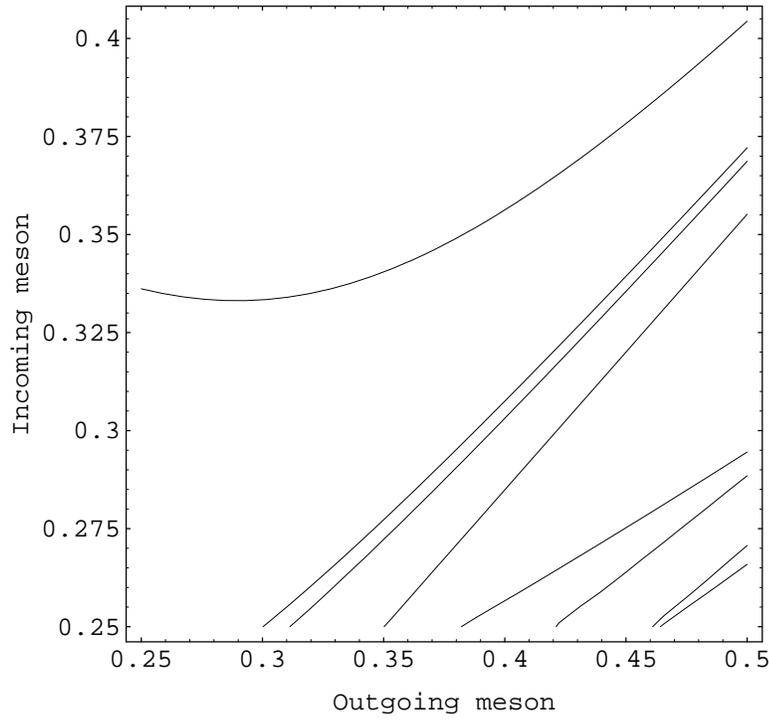}
\caption{Nodal lines of $3\: ^1S_0\; \pi(1800)\rightarrow\rho\pi,\; \rho\omega,\; 
K^* K,\;  K^* K^*$ as a function of the incoming and outgoing meson
$\beta_A$ and $\beta$ in GeV. For each channel there are two lines of nodes.
From top to bottom the nodal lines correspond to the 
$\rho\pi,\; \rho\omega,\;  K^*K,\;  K^* K^*,\;  \rho\pi,\;  K^*K,\;  \rho\omega,\;  K^* K^*$ channels. }
\plabel{3snodes}
\end{figure}

% Our analysis in the previous section or data on
%the $\rho(1450)$ suggest that $\rho(1450)$ feeds $4\pi$ and so both
%$\pi \rho(2S)$ and $\rho \omega$ should
%give a prominent signal for $3\: ^1S_0\;\pi \rightarrow 5\pi$. The data on $\pi(1800)$
%are not like this \cite{khok}. ***(5 pi spectrum peaks in some circumstances
%at 1725 MeV at 300 MeV [qu nos not known].)***

\section{Summary and Experimental  Strategy}

The $\rho(1450)$ and $\omega(1420)$ have masses that are consistent
with radial $2S$ but their decays have a strong hybrid character, as already
noted \cite{cd1,cp95}. We find that
both of these states and the heavier counterpart 
$\omega(1600)$ can be interpreted as $2S$--hybrid mixtures. Present data
on the $\rho(1700)$ are consistent with it being a $2S$--hybrid mixture but
do not demand it. 
We note that three $\rho$ (and three $\omega$) states
should be allowed for between $1300 - 1800$ MeV in future data
analyses. 

The $3\: ^1S_0$ $\pi$ is expected in the $1800$ MeV mass region as is the
hybrid. We find that the decay patterns of these are very different.
The low total width state with strong $\pi f_0$ (hybrid) and the large
total width state with strong $\rho \omega$ ($3S$) is the sharpest
discriminant. The established VES state $\pi(1800)$ clearly exhibits
the former hybrid character. We also urge data analysts to allow for the possibility of two isovector
$0^{-+}$ resonances in the region $1700 - 1900$ MeV, one of which is
expected to couple strongly to $\rho \omega$.

\vspace{.5cm}

We are indebted to D.V.Bugg, S.-U.Chung, A.Donnachie, A.Kirk, 
I.Kachaev, Y.A.Khokhlov, D.I.Ryabchikov and A.M.Zaitsev for
discussions. FEC is partially supported by the European Community
Human Mobility Program Eurodafne, Contract CHRX-CT92-0026.

\end{document}